\documentclass[10pt]{article}

\usepackage{bm,subfigure,graphicx,cite}

\pdfoutput=1

\graphicspath{{./images/}}

\author{Nasser Mohieddin Abukhdeir and Alejandro D. Rey\\
nasser.abukhdeir@mcgill.ca, alejandro.rey@mcgill.ca\\
Department of Chemical Engineering\\
McGill University\\
Montr\'{e}al, Qu\'{e}bec, Canada}

\title{Non-isothermal model for the direct isotropic/smectic-A liquid crystalline transition}

\begin{document}

\maketitle

\abstract{An extension to a high-order model for the direct isotropic/smectic-A liquid crystalline phase transition was derived to take into account thermal effects including anisotropic thermal diffusion and latent heat of phase-ordering. Multi-scale multi-transport simulations of the non-isothermal model were compared to isothermal simulation, showing that the presented model extension corrects the standard Landau-de Gennes prediction from constant growth to diffusion-limited growth, under shallow quench/undercooling conditions. Non-isothermal simulations, where meta-stable nematic pre-ordering precedes smectic-A growth, were also conducted and novel non-monotonic phase-transformation kinetics observed.}

\section{INTRODUCTION} \label{sec:intro}

\begin{figure} 
\centering
\includegraphics[width=2.5in]{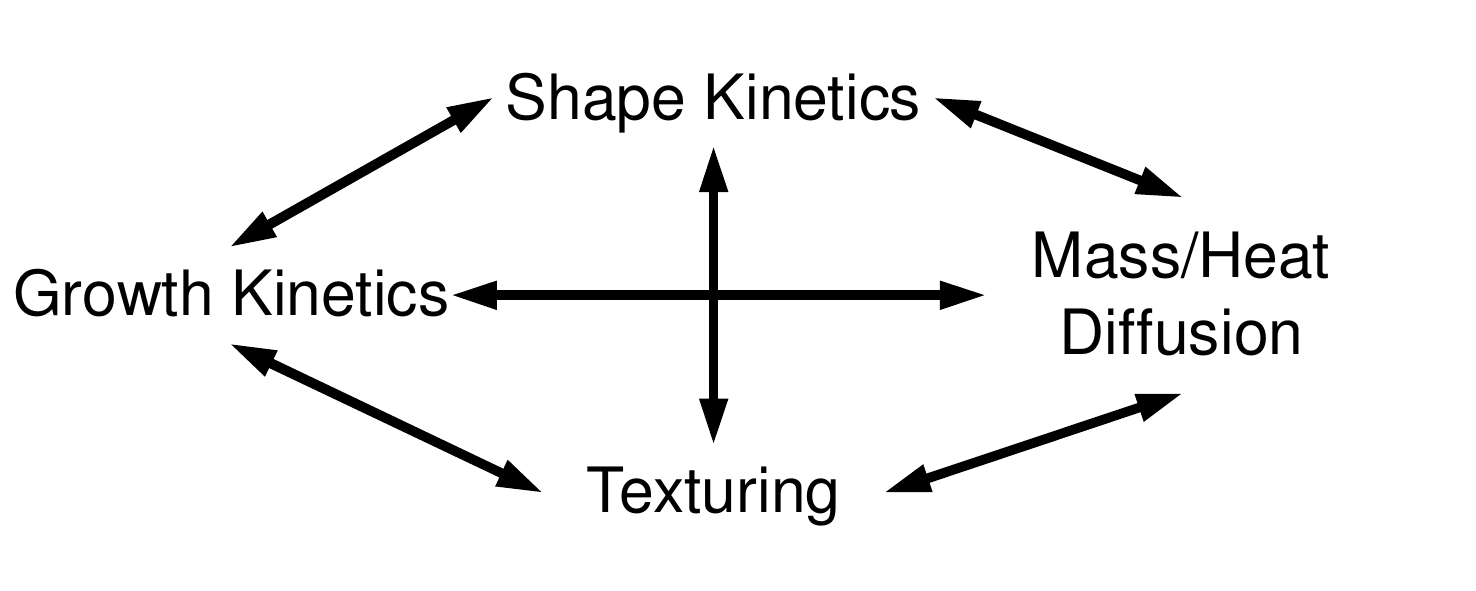}
\caption{Schematic showing the inter-relationships between the phase-ordering processes. Adapted from Figure 1 of ref. \cite{Abukhdeir2008b}}
\label{fig:summary}
\end{figure}

The kinetics of phase transformations \cite{Balluffi2005} is a fundamental subject in material and interfacial science that has widespread impact on material manufacturing and use. Three fundamental kinetic phenomena associated with phase transformations are microstructure, growth rate, and shape evolution \cite{Sutton1995} (see \ref{fig:summary}). The growth processes associated with phase transformations vary, depending on driving forces, such as bulk free energy minimization and reduction of interfacial area. Growth laws in diffusive and non-diffusive transformations typically differ in that the former has a conserved order parameter \cite{Balluffi2005,Sutton1995}. The differing kinetics of non-diffusive and diffusive phase transitions are manifested in their growth laws $l \propto t^n$, where $l$ is the characteristic length of the growing domain. Non-diffusive phase transitions exhibit constant growth where $l$ scales linearly  ($n=1$). Diffusive phase transitions exhibit diffusion-limited growth, where $n=\frac{1}{2}$.

In the case of liquid crystals, both (mass and thermal) diffusive and (phase-ordering) non-diffusive transformation dynamics are present simultaneously. The study of diffusive transformations in liquid crystalline materials has mainly focused on mass diffusion. Two main areas of this research involve liquid crystalline materials where either a small amount of (undesired) impurity is present or where a composite material is desired, such as polymer dispersed liquid crystals \cite{Drzaic1995} (PDLCs). In the former case, much work has focused on the study of diffusion-driven growth instabilities \cite{Mullins1963,Mullins1964} (see refs. \cite{Coriell1985,Bechhoefer1996} and Chapter B.IX of ref. \cite{Oswald2005}) in nematic liquid crystals both through experimental and theoretical approaches \cite{Bechhoefer1989,Simon1990,Oswald1991,Bechhoefer1995,Misbah1995,Mesquita1996,Ignes-Mullol2000,Gomes2001}. Higher order mesophases such as smectic and columnar liquid crystals have been studied, in this context, to a lesser extent \cite{Gonzalez-Cinca1998,Oswald2006}. Much of this interest stems from the fact that liquid crystal transformations occur on experimentally accessible timescales and are inherently anisotropic. In the latter case, the study of PDLC materials \cite{Drzaic1995} has focused on different mechanisms of phase-separation \cite{Dorgan1993,Amundson1998,Borrajo1998,Lucchetti2000,Harrison2000a,Nakazawa2001,Hoppe2002,Hoppe2004,Hoppe2004a,Das2006} (polymerization, solvent, and thermally induced), where growth morphology of liquid crystal-rich domains and their texture are studied in order to control functional physical properties, such as electro-optical response. Subsequently, PDLCs composed of higher order liquid crystals have also been studied to a lesser extent \cite{Benmouna1999,Benmouna2000,Graca2003}.

On the other hand, thermal diffusion in liquid crystalline phase transitions has typically been neglected due to the relatively low latent heat contributions of (nematic) liquid crystalline transformations and the pervasive use of thin film geometries \cite{Bechhoefer1996} (where evolved latent heat escapes in the vertical dimension). Focusing on pure/single-phase liquid crystals, where impurity concentrations are below the saturation limit, evidence has shown that latent heat effects can also result in diffusive dynamics \cite{Huisman2007}. Thus heat diffusion must be taken into account to reproduce experimentally observed growth laws under these conditions \cite{Dierking2001,Dierking2003}. As the study of these non-isothermal effects progresses \cite{Huisman2007,Abukhdeir2008b,Soule2008}, it is becoming clear that they play a non-negligible role in the complex kinetics and dynamics of liquid crystalline phase transitions.

As previously mentioned, a relatively large amount of effort in the study of liquid crystals has focused on those which exhibit only some degree of orientational order, or nematics. Higher-order mesophases that exhibit some degree of positional order, in addition to orientational, have been less studied. The most simple of these higher order liquid crystals is the smectic-A mesophase, which exhibits lamellar translational ordering, in addition to the orientational ordering of nematics.  Recently, an increasing amount of interest in this mesophase, in particular of materials exhibiting a direct isotropic/smectic-A (disordered/ordered) transition, has resulted in many experimental and theoretical results.  Nonetheless, the understanding of this mesophase is in a nascent stage.  Much of this is due to the time and length scales at which the structures and dynamics occur being on the nano-scale.

The additional positional order of smectics, where the underlying primary order is orientational, allows the possibility of dual-order kinetics and dynamics. This has been experimentally observed in a smectic polymer liquid crystal which exhibits extremely slow dynamics \cite{Tokita2006}. The presence of a transient meta-stable nematic phase preceding the stable smectic phase in this polymer liquid crystal was observed under certain quench conditions \cite{Tokita2004}. The general phenomena of meta-stable states in phase transitions involving dual non-conserved order was theoretically shown over a decade ago \cite{Bechhoefer1991}, but this generalized approach is not suitable for liquid crystal phase transitions (\ref{fig:summary}). A first-approximation of the experimental system \cite{Tokita2006}, taking into account shape kinetics/growth kinetics/texturing (see \ref{fig:summary}), has been modeled \cite{Abukhdeir2009} shedding light on some of the nano-scale phenomena resulting in these experimental observations.

In this context, the kinetics of liquid crystal phase transitions, in particular smectics, becomes more complex and diverse than currently regarded where coupled non-diffusive (phase-ordering), diffusive (mass/thermal), and pre-ordering transformation dynamics are present. The general objective of this work is to contribute to the fundamental understanding of mesophase formation under simultaneous orientational and positional symmetry-breaking and under non-negligible latent heat evolution. The specific objectives of this work are (1) to extend an existing high-order isothermal model for the direct isotropic/smectic-A transition to account for thermal effects and (2) to conduct a basic study of the phase transition landscape:
\begin{enumerate}
\item Derive an energy balance to extend an existing high-order model for the isotropic/smectic-A phase transition \cite{deGennes1995,Mukherjee2001} to account for non-isothermal effects including latent heat of phase-ordering and thermal diffusion.
\item Simulate a one-dimensional growing smectic-A front, using the extended model, to show that the non-isothermal extension corrects the standard Landau-de Gennes prediction of volume-driven growth kinetics ($l \propto t$) to diffusion-limited growth kinetics ($l \propto t^{1/2}$) under shallow quench/undercooling conditions, consistent with experimental observations \cite{Dierking2001,Dierking2003,Oswald2006}.
\item Simulate a one-dimensional growing smectic-A front using the non-isothermal model under conditions where meta-stable nematic pre-ordering is observed \cite{Abukhdeir2009a}.
\end{enumerate}
This work neglects nucleation mechanisms \cite{Ziabicki2003}, fluctuations \cite{Singh2002,Pleiner1996}, impurities \cite{Drzaic1995,Oswald2005}, and convective flow \cite{Pleiner1996} while taking into account energetically the inter-coupling between orientational/translational order and variation of smectic layer spacing.  This work is organized into effectively three sections: background/model, energy balance derivation, and simulation. The background/model section provides a brief introduction to relevant types of liquid crystal phase-ordering and the Landau-de Gennes type model of Mukherjee, Pleiner, and Brand \cite{deGennes1995,Mukherjee2001} used in this work. The derivation of the thermal energy balance extension to this model is then derived. Finally, simulation results are presented and discussed, followed by a summary of the conclusions.

\section{BACKGROUND AND MODEL}

\subsection{LIQUID CRYSTAL ORDER}

As mentioned above, liquid crystalline phases or mesophases are materials which exhibit partial orientational and/or translational order.  They are composed of anisotropic molecules  which can be disc-like (discotic) or rod-like (calamitic) in shape.  Thermotropic liquid crystals are typically pure-component compounds that exhibit mesophase ordering most greatly in response to temperature changes.  Lyotropic liquid crystals are mixtures of mesogens (molecules which exhibit some form of liquid crystallinity), possibly with a solvent, that most greatly exhibit mesophase behavior in response to concentration changes.  Effects of pressure and external fields also influence mesophase behavior.  This work focuses on the study of calamitic thermotropic liquid crystals which exhibit a first-order mesophase transition.

\begin{figure}
\centering
\subfigure[]{\includegraphics[width=0.8in]{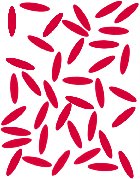}}
\subfigure[]{\includegraphics[width=0.8in]{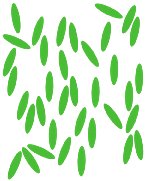}}
\subfigure[]{\includegraphics[width=0.8in]{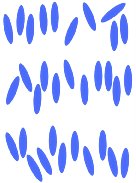}}
\caption{schematics of the (a) isotropic, (b) nematic, and (c) smectic-A phases.}
\label{fig:lcorder}
\end{figure}

An unordered liquid, where there is neither orientational nor translational order (apart from an average intermolecular separation distance) of the molecules, is referred to as isotropic. Liquid crystalline order involves partial orientational order (nematics) and, additionally, partial translational order (smectics and columnar mesophases).  The simplest of the smectics is the smectic-A mesophase, which exhibits one-dimensional translational order in the direction of the preferred molecular orientational axis.  It can be thought of as layers of two-dimensional fluids stacked upon each other.  Schematic representation of these different types of ordering is shown in \ref{fig:lcorder}.

Theoretical characterization of orientational and translational order of the smectic-A mesophase (see \ref{fig:lcorder}) is accomplished using order parameters that adequately capture the physics involved.   Partial orientational order of the nematic phase is characterized using a symmetric traceless quadrupolar tensor \cite{deGennes1995}:\begin{equation} \label{eqn:nem_order_param}
\bm{Q} = S \left(\bm{nn} - \frac{1}{3} \bm{I}\right) + \frac{1}{3} P \left( \bm{mm} - \bm{ll}\right)
\end{equation}                   
where $\mathbf{n}/\mathbf{m}/\mathbf{l}$ are the eigenvectors of $\bm{Q}$, which characterize the average molecular orientational axes, and $S/P$ are scalars which characterize the extent to which the molecules conform to the average orientational axes \cite{Rey2002,Yan2002,Rey2007}.  Uniaxial order is characterized by $S$ and $\bm{n}$, which correspond to the maximum eigenvalue (and its corresponding eigenvector) of $\bm{Q}$, $S= \frac{3}{2} \mu_n$.  Biaxial order is characterized by $P$ and $\bm{m}/\bm{l}$, which correspond to the lesser eigenvalues and eigenvectors, $P = \frac{3}{2}\left(\mu_m - \mu_l\right)$.

The one-dimensional translational order of the smectic-A mesophase in addition to the orientational order found in nematics is characterized through the use of primary (orientational) and secondary (translational) order parameters together \cite{Toledano1987}.  A complex order parameter can be used to characterize translational order \cite{deGennes1995}:
\begin{equation} \label{eqsmec_order_param}
\Psi = \psi e^{i \phi} = A+iB
\end{equation}
where $\phi$ is the phase, $\psi$ is the scalar amplitude of the density modulation, and $A/B$ is the real/imaginary component of the complex order parameter. The density wave vector, which describes the average orientation of the smectic-A density modulation, is defined as $\mathbf{a} = \nabla \phi / |{\nabla \phi}|$.  The smectic scalar order parameter $\psi$ characterizes the magnitude of the density modulation and is used in a dimensionless form in this work.  In the smectic-A mesophase the preferred orientation of the wave vector is parallel to the average molecular orientational axis, $\mathbf{n}$.

\subsection{MODEL FOR THE DIRECT ISOTROPIC/SMECTIC-A TRANSITION}

A two-order parameter Landau-de Gennes model for the first order isotropic/smectic-A phase transition is used that was initially presented by Mukherjee, Pleiner, and Brand \cite{deGennes1995,Mukherjee2001} and later extended by adding nematic elastic terms \cite{Brand2001,Mukherjee2002a}:
\begin{eqnarray} \label{eq:free_energy_heterogeneous}
f - f_0 =&\frac{1}{2} a \left(\bm{Q} : \bm{Q}\right) - \frac{1}{3} b \left(\bm{Q}\cdot\bm{Q}\right) : \bm{Q} + \frac{1}{4} c \left(\bm{Q} : \bm{Q}\right)^2   \nonumber\\
& + \frac{1}{2} \alpha \left|\Psi\right|^2 + \frac{1}{4} \beta \left|\Psi\right|^4 \nonumber\\
&- \frac{1}{2} \delta \left| \Psi \right|^2 \left(\bm{Q} : \bm{Q}\right) - \frac{1}{2} e \bm{Q}:\left(\bm{\nabla} \Psi\right)\left(\bm{\nabla} \Psi^*\right) \nonumber\\
& + \frac{1}{2} l_1 \left(\bm{\nabla} \bm{Q} \right)^2 + \frac{1}{2} l_2 \left( \bm{\nabla} \cdot \bm{Q} \right)^2 \nonumber\\
& + \frac{1}{2} l_3 \bm{Q}:\left( \nabla \bm{Q} : \nabla \bm{Q} \right) \nonumber\\ 
&+ \frac{1}{2} b_1 \left|\bm{\nabla} \Psi\right|^2 + \frac{1}{4} b_2 \left|\nabla^2 \Psi\right|^2
\end{eqnarray}
\begin{equation} \label{eq:free_energy_heterogenous_coeffs}
a =  a_0 (T - T_{NI}) ; \alpha = \alpha_0 (T - T_{AI})\nonumber 
\end{equation}
where $f$ is the free energy density, $f_0$ is the free energy density of the isotropic phase, terms 1-5 are the bulk contributions to the free energy, terms 6-7 are couplings of nematic and smectic order; both the bulk order and coupling of the nematic director and smectic density-wave vector, respectively.  Terms 8-10/11-12 are the nematic/smectic elastic contributions to the free energy.  $T$ is temperature, $T_{NI}$/$T_{AI}$ are the hypothetical second order transition temperatures for isotropic/nematic and isotropic/smectic-A mesophase transitions (refer to \cite{Coles1979a} for more detail), and the remaining constants are phenomenological parameters. This free energy density expression is a real-valued function of the complex order parameter $\Psi$ and its complex conjugate $\Psi^*$, which makes it convenient to reformulate using the real and imaginary parts of the complex order parameter (see eqn \ref{eqsmec_order_param}).

\begin{figure}
\centering
\includegraphics[width=2.5in]{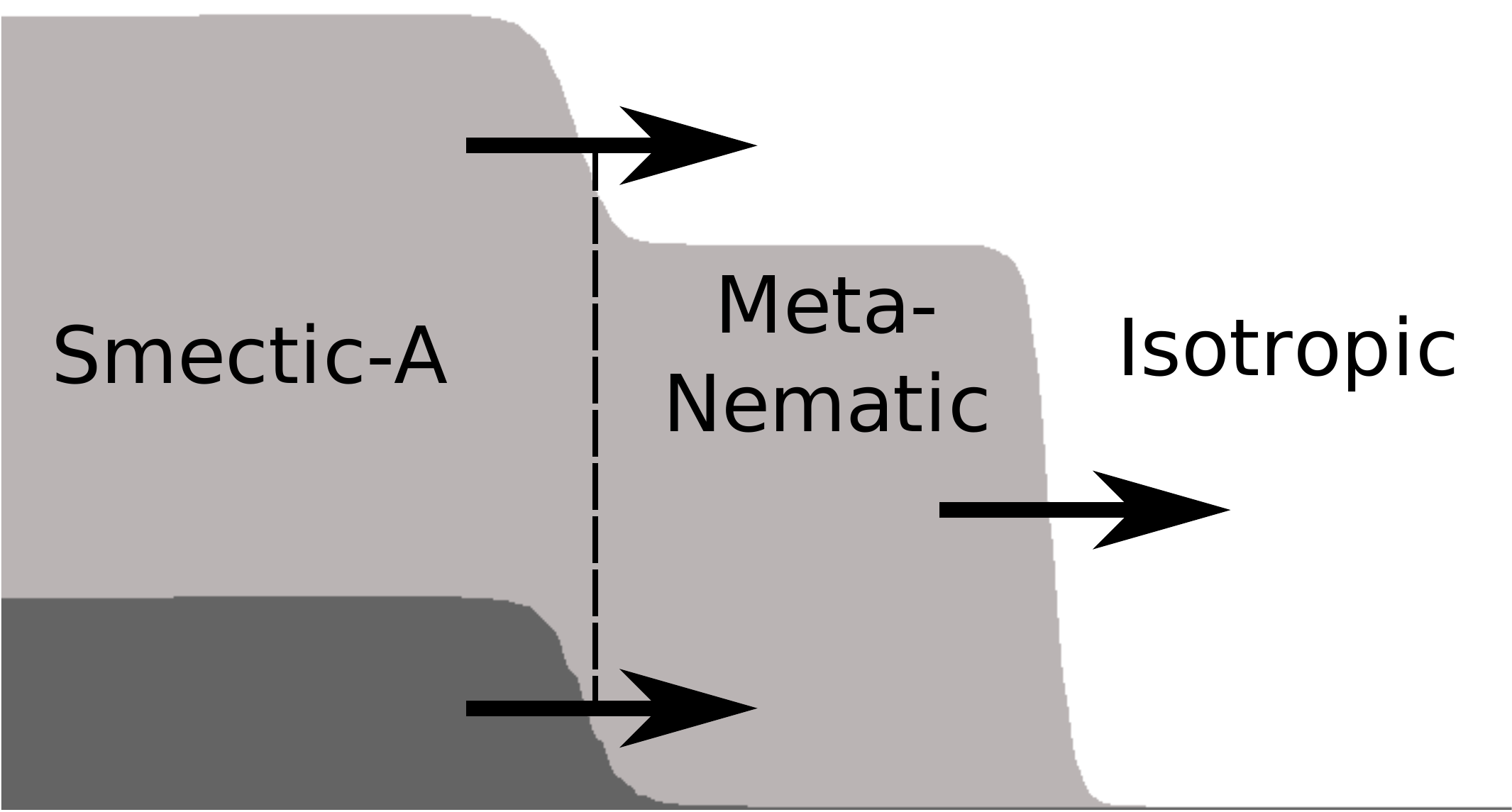}
\caption{A schematic of one-dimensional direct isotropic/smectic-A growth under conditions where meta-stable nematic pre-ordering is present. The light/dark gray shading indication the magnitude of the nematic/smectic-A scalar order parameter $S$/$\psi$ growing in an unstable isotropic phase. Arrows indicate growth direction, where the meta-stable nematic front grows independently from the trailing smectic-A front. The increase in the nematic scalar order parameter is due to coupled enhancement from the presence of smectic order \cite{Coles1979,Coles1979a,Mukherjee2001}.}
\label{fig:meta}
\end{figure}

As previously mentioned, the dual-order nature of the isotropic/smectic-A liquid crystal transition has been shown to result in meta-stable nematic pre-ordering. \ref{fig:meta} shows a schematic of this phenomena, where a meta-stable nematic front precedes the stable smectic-A front, growing into an isotropic matrix phase. Under standard conditions, meta-stable nematic pre-ordering is not observed, but when the dynamic timescales between nematic/orientational and smectic-A/translational ordering differ greatly, meta-stable nematic pre-ordering is observed and predicted by the model based on eqn. \ref{eq:free_energy_heterogeneous} \cite{Bechhoefer1991,Tuckerman1992,Tokita2006,Abukhdeir2009a}.

The previously mentioned dynamic timescales are included in this model via the Landau-Ginzburg time-dependent formulation \cite{Barbero2000}. The general form of the time-dependent formulation is as follows \cite{Barbero2000}:
\begin{eqnarray} \label{eq:landau_ginz}
\left(\begin{array}{c}
 \frac{\partial \bm{Q}}{\partial t}
\\ \frac{\partial A}{\partial t}
\\ \frac{\partial B}{\partial t} 
\end{array}\right)
&=& 
\left(\begin{array}{c c c} 
\frac{1}{\mu_n} & 0 & 0\\ 
0 & \frac{1}{\mu_S} & 0\\ 
0 & 0 & \frac{1}{\mu_S}\end{array} \right)
\left(\begin{array}{c} -\frac{\delta F}{\delta \bm{Q}}\\ 
-\frac{\delta F}{\delta A}\\ 
-\frac{\delta F}{\delta B} \end{array}\right)\\
F &=& \int_V f dV
\end{eqnarray}
where $\mu_n$/$\mu_s$ is the rotational/smectic viscosity, and $V$ the control volume. A higher order functional derivative must be used due to the second-derivative term in the free energy equation (\ref{eq:free_energy_heterogeneous}):
\begin{equation} \label{eqpdes}
\frac{\delta F}{\delta \theta} = \frac{\partial f}{\partial \theta} - \frac{\partial}{\partial x_i}\left(\frac{\partial f}{\partial \frac{\partial \theta}{\partial x_i}} \right) + \frac{\partial}{\partial x_i}\frac{\partial}{\partial x_j}\left(\frac{\partial f}{\partial \frac{\partial^2 \theta}{\partial x_i \partial x_j}} \right)
\end{equation}
where $\theta$ corresponds to the order parameter.

\subsection{THERMAL ENERGY BALANCE DERIVATION}

The energy balance for a differential volume, without flow, is:
\begin{equation} \label{eqn:govn}
\frac{d u}{d t} = \frac{d W}{d t} - \nabla \cdot \bm{q}
\end{equation}
where $u$ is the internal energy density, $\bm{q}$ is the heat flux, and $\frac{d W}{d t}$ is the rate of mechanical work on the differential control volume. According to eq. 4.46 of ref \cite{Stewart2004}, the rate of mechanical work on the system can be written as the sum of elastic energy stored and the rate of dissipation. The rate of elastic energy storage is the change in nematic/smectic-A energy due to changes in the order parameters, so it is $\frac{\partial f}{\partial t}$ taken at constant temperature. The rate of dissipation in the absence of flow is equal to \cite{Qian1998}:
\begin{equation}
\bm{H}_{\bm{Q}}:\frac{\partial \bm{Q}}{\partial t} + H_A\frac{\partial A}{\partial t}  + H_B \frac{\partial B}{\partial t} 
\end{equation}
where $A/B$ is the real/imaginary part of the complex order parameter ($\Psi$), the molecular fields $\bm{H}_{\bm{Q}},H_A,H_B$ are the negative of the variational derivatives of the free energy functional with respect to $\bm{Q},A,B$. With no flow, the Landau-de Gennes model gives (symmetric-traceless contributions are implied throughout this work):
\begin{eqnarray}
\bm{H}_{\bm{Q}} = \mu_n \frac{\partial \bm{Q}}{\partial t} = -\frac{\partial f}{\partial \bm{Q}} + \bm{\nabla} \cdot \left(\frac{\partial f}{\partial \bm{\nabla} \bm{Q}} \right) \nonumber\\
H_{A} = \mu_s \frac{\partial A}{\partial t} = -\frac{\partial f}{\partial A} + \bm{\nabla} \cdot \left(\frac{\partial f}{\partial \bm{\nabla} A} \right) - \bm{\nabla \nabla} : \left(\frac{\partial f}{\partial \bm{\nabla \nabla} A} \right) \nonumber\\
H_{B} = \mu_s \frac{\partial B}{\partial t} = -\frac{\partial f}{\partial B} + \bm{\nabla} \cdot \left(\frac{\partial f}{\partial \bm{\nabla} B} \right) - \bm{\nabla \nabla} : \left(\frac{\partial f}{\partial \bm{\nabla \nabla} B} \right)
\end{eqnarray}
The rate of mechanical work is thus:
\begin{eqnarray}
\frac{d W}{d t} &=& \frac{\partial f}{\partial \bm{Q}} : \frac{\partial \bm{Q}}{\partial t} + \frac{\partial f}{\partial \bm{\nabla} \bm{Q}} \vdots \frac{\partial \bm{\nabla} \bm{Q}}{\partial t} \nonumber\\
&+& \frac{\partial f}{\partial A} \frac{\partial A}{\partial t} + \frac{\partial f}{\partial \bm{\nabla} A} \cdot \frac{\partial \bm{\nabla} A}{\partial t} + \frac{\partial f}{\partial \bm{\nabla\nabla} A} : \frac{\partial \bm{\nabla\nabla} A}{\partial t} \nonumber\\
&+& \frac{\partial f}{\partial B} \frac{\partial B}{\partial t} + \frac{\partial f}{\partial \bm{\nabla} B} \cdot \frac{\partial \bm{\nabla} B}{\partial t} + \frac{\partial f}{\partial \bm{\nabla\nabla} B} : \frac{\partial \bm{\nabla\nabla} B}{\partial t} \nonumber\\
&+& \bm{H}_{\bm{Q}}:\frac{\partial \bm{Q}}{\partial t} + H_A\frac{\partial A}{\partial t}  + H_B \frac{\partial B}{\partial t} 
\end{eqnarray}
which simplifies to:
\begin{eqnarray}\label{eqn:final_mech}
\frac{d W}{d t} &=& \frac{\partial f}{\partial \bm{\nabla} \bm{Q}} \vdots \frac{\partial \bm{\nabla} \bm{Q}}{\partial t} \nonumber\\
&+& \frac{\partial f}{\partial \bm{\nabla} A} \cdot \frac{\partial \bm{\nabla} A}{\partial t} + \frac{\partial f}{\partial \bm{\nabla\nabla} A} : \frac{\partial \bm{\nabla\nabla} A}{\partial t} \nonumber\\
&+& \frac{\partial f}{\partial \bm{\nabla} B} \cdot \frac{\partial \bm{\nabla} B}{\partial t} + \frac{\partial f}{\partial \bm{\nabla\nabla} B} : \frac{\partial \bm{\nabla\nabla} B}{\partial t} \nonumber\\
&+& \left( \bm{\nabla} \cdot \left(\frac{\partial f}{\partial \bm{\nabla} \bm{Q}} \right) \right):\frac{\partial \bm{Q}}{\partial t}\nonumber\\ 
&+& \left( \bm{\nabla} \cdot \left(\frac{\partial f}{\partial \bm{\nabla} A} \right) - \bm{\nabla \nabla} : \left(\frac{\partial f}{\partial \bm{\nabla \nabla} A} \right) \right)\frac{\partial A}{\partial t} \nonumber\\
&+& \left( \bm{\nabla} \cdot \left(\frac{\partial f}{\partial \bm{\nabla} B} \right) - \bm{\nabla \nabla} : \left(\frac{\partial f}{\partial \bm{\nabla \nabla} B} \right) \right)\frac{\partial B}{\partial t} 
\end{eqnarray}
The left-hand side of eq \ref{eqn:govn} can be written in terms of Helmholtz free energy:
\begin{equation} \label{eqn:int_energy}
\frac{d u}{d t} = \frac{\partial}{\partial t} \left( f - T \frac{\partial f}{\partial T} \right )= \frac{\partial f}{\partial t} -  \frac{\partial f}{\partial T} \frac{\partial T}{\partial t} - T \frac{\partial^2 f}{ \partial t \partial T}
\end{equation}
expanding the time derivatives yields:
\begin{eqnarray}
\frac{\partial f}{\partial t} & = & \frac{\partial f}{\partial T}\frac{\partial T}{\partial t} +\frac{\partial f}{\partial \bm{Q}} : \frac{\partial \bm{Q}}{\partial t} + \frac{\partial f}{\partial \bm{\nabla} \bm{Q}} \vdots \frac{\partial \bm{\nabla} \bm{Q}}{\partial t} \nonumber\\
&+& \frac{\partial f}{\partial A} \frac{\partial A}{\partial t} + \frac{\partial f}{\partial \bm{\nabla} A} \cdot \frac{\partial \bm{\nabla} A}{\partial t} + \frac{\partial f}{\partial \bm{\nabla\nabla} A} : \frac{\partial \bm{\nabla\nabla} A}{\partial t} \nonumber\\
&+& \frac{\partial f}{\partial B} \frac{\partial B}{\partial t} + \frac{\partial f}{\partial \bm{\nabla} B} \cdot \frac{\partial \bm{\nabla} B}{\partial t} + \frac{\partial f}{\partial \bm{\nabla\nabla} B} : \frac{\partial \bm{\nabla\nabla} B}{\partial t} \label{eqn:1}\\
- T  \frac{\partial^2 f}{\partial t \partial T} & =&  -T \frac{\partial^2 f}{\partial T^2} \frac{\partial T}{\partial t}\nonumber \\
&-& T \left( \frac{\partial^2 f}{\partial \bm{Q} \partial T}:\frac{\partial \bm{Q}}{\partial t} + \frac{\partial^2 f}{\partial A \partial T} \frac{\partial A}{\partial t} + \frac{\partial^2 f}{\partial B \partial T}\frac{\partial B}{\partial t} \right) \label{eqn:2}
\end{eqnarray}
where it has been taken into account in eqn. \ref{eqn:2} that all gradient terms in the free energy $f$ are assumed temperature-independent. Additionally, the first term in eqn. \ref{eqn:2} includes the specific heat (per unit volume):
\begin{equation}\label{eqn:Cp}
C_p = - T \frac{\partial^2 f}{\partial T^2}
\end{equation}
Substituting eqns. \ref{eqn:1}-\ref{eqn:Cp} into eqn. \ref{eqn:int_energy} yields the total change of internal energy:
\begin{eqnarray}\label{eqn:final_int_energy}
\frac{d u}{d t} &=& C_p \frac{\partial T}{\partial t} - T \left( \frac{\partial^2 f}{\partial \bm{Q} \partial T}:\frac{\partial \bm{Q}}{\partial t} + \frac{\partial^2 f}{\partial A \partial T} \frac{\partial A}{\partial t} + \frac{\partial^2 f}{\partial B \partial T}\frac{\partial B}{\partial t} \right) \nonumber\\
&+&\frac{\partial f}{\partial \bm{Q}} : \frac{\partial \bm{Q}}{\partial t} + \frac{\partial f}{\partial \bm{\nabla} \bm{Q}} \vdots \frac{\partial \bm{\nabla} \bm{Q}}{\partial t} \nonumber\\
&+& \frac{\partial f}{\partial A} \frac{\partial A}{\partial t} + \frac{\partial f}{\partial \bm{\nabla} A} \cdot \frac{\partial \bm{\nabla} A}{\partial t} + \frac{\partial f}{\partial \bm{\nabla\nabla} A} : \frac{\partial \bm{\nabla\nabla} A}{\partial t} \nonumber\\
&+& \frac{\partial f}{\partial B} \frac{\partial B}{\partial t} + \frac{\partial f}{\partial \bm{\nabla} B} \cdot \frac{\partial \bm{\nabla} B}{\partial t} + \frac{\partial f}{\partial \bm{\nabla\nabla} B} : \frac{\partial \bm{\nabla\nabla} B}{\partial t} 
\end{eqnarray}
Finally, substituting the final expression for the total change in internal energy eqn. \ref{eqn:final_int_energy} and rate of mechanical work eqn. \ref{eqn:final_mech} into the energy balance eqn. \ref{eqn:govn} yields:
\begin{eqnarray} \label{eqn:final_energy_balance}
C_p \frac{\partial T}{\partial t} &=& \mu_n \frac{\partial \bm{Q}}{\partial t} : \frac{\partial \bm{Q}}{\partial t} + \mu_s \left( \frac{\partial A}{\partial t} : \frac{\partial A}{\partial t} + \frac{\partial B}{\partial t} : \frac{\partial B}{\partial t} \right) \nonumber\\
&+&  T \left( \frac{\partial^2 f}{\partial \bm{Q} \partial T}:\frac{\partial \bm{Q}}{\partial t} + \frac{\partial^2 f}{\partial A \partial T} \frac{\partial A}{\partial t} + \frac{\partial^2 f}{\partial B \partial T}\frac{\partial B}{\partial t} \right) \nonumber\\
&-& \bm{\nabla} \cdot \bm{q}
\end{eqnarray}
This can be further simplified taking the derivatives of free energy density eqn. \ref{eq:free_energy_heterogeneous}:
\begin{eqnarray}
C_p \frac{\partial T}{\partial t} &=& \mu_n \frac{\partial \bm{Q}}{\partial t} : \frac{\partial \bm{Q}}{\partial t} + \mu_s \left( \frac{\partial A}{\partial t} : \frac{\partial A}{\partial t} + \frac{\partial B}{\partial t} : \frac{\partial B}{\partial t} \right) \nonumber\\
&+&  T \left( \frac{1}{2} a_0 \bm{Q}:\frac{\partial \bm{Q}}{\partial t} + \frac{1}{2} \alpha_0 \left( A \frac{\partial A}{\partial t} + B \frac{\partial B}{\partial t} \right) \right) \nonumber\\
&-& \bm{\nabla} \cdot \bm{q}
\end{eqnarray}
where term 1/2 corresponds to orientational/translational dissipation, term 3 corresponds to the energy of phase-ordering, and term 4 is the heat flux. A constitutive relationship for the heat flux that can be used is the anisotropic Fourier's law:
\begin{equation} \label{eq:fourier}
\bm{q} = - \bm{K}\cdot \nabla T   
\end{equation}                                                     
where the thermal conductivity tensor $\bm{K}$ is used due to the anisotropy of the nematic phase. The thermal conductivity tensor can be written as the sum of isotropic and anisotropic contributions:
\begin{equation} \label{eq:thermal_conductivity}
\bm{K} = k_{iso} \bm{\delta} + k_{an}\bm{Q} = \left(\frac{k_{\parallel}+2 k_{\perp}}{3}\right) \bm{\delta}+ \left(k_{\parallel}- k_{\perp}\right)\bm{Q}
\end{equation}   
where $k_{iso}$ and $k_{an}$ are the isotropic and anisotropic contributions to the thermal conductivities, and $k_{\parallel}$ and $k_{\perp}$ are the conductivities in the directions parallel and perpendicular to the average orientational axis (nematic director), respectively. While eqn \ref{eq:thermal_conductivity} takes into account the dependence of thermal conductivity on the degree of orientational order \cite{Zammit1990,Marinelli1996}, the effect of the translation ordering of the smectic-A phase on the thermal conductivity is neglected based upon past experimental work. This work determined that the shape anisotropy of the mesogens (liquid crystal molecules) is the main contribution to the thermal transport anisotropy compared to smectic layering \cite{Rondelez1978}. Eqns. \ref{eq:free_energy_heterogeneous},\ref{eq:landau_ginz},\ref{eqn:final_energy_balance},\ref{eq:thermal_conductivity} constitute the model system of equations.

\subsection{SIMULATION CONDITIONS} \label{sec:sim}

One-dimensional simulation of the model was performed using the Galerkin finite element method (Comsol Multiphysics). Quadratic Lagrange basis functions were used for the Q-tensor variables/temperature and quartic Hermite basis functions used for the complex order parameter components. Standard numerical techniques were utilized to ensure convergence and stability of the solution including an adaptive backward-difference formula implicit time integration method. A uniform mesh was used such that there was a density of $6$ nodes per equilibrium smectic-A layer in the liquid crystal computational domain of $500$ layers (approximately $2 \mu m$). A coupled thermal domain, where only transient thermal diffusion was solved for, leveraged the fact that the characteristic thermal length scale (order of $\mu m$) is much greater than that of the characteristic smectic-A/meta-stable nematic length scale \cite{Abukhdeir2008c} (order of $nm$). Neumann boundary conditions were used at both boundaries and the initial condition consisted of uniform temperature with a single smectic-A layer nucleus. Additionally, a correction factor $\Theta=0.96$ was used to normalize the slight difference between the latent heat value predicted by the phenomenological model parameters used (term 3 of eqn. \ref{eqn:final_energy_balance}) to an experimentally determined approximate reference value of $\Delta H_{exp} = 4.85 kJ/mol$ \cite{Oweimreen2000}. The latent heat predicted by the model and phenomenological parameters (determined previously \cite{Abukhdeir2008c}) is:
\begin{equation}
\Delta H = T_b \frac{\partial f (T_b)}{\partial T} = T_b \left(\frac{1}{3} a_0 S_{b}^2 + \frac{1}{2} \alpha_0 \psi_{b}^2\right)
\end{equation}
where $T_b$ is the bulk isotropic/smectic-A transition temperature, based upon an approximate reference value experimentally determined \cite{Coles1979a} to be $331.35 K$.

\begin{figure} 
\centering
\includegraphics[width=3in]{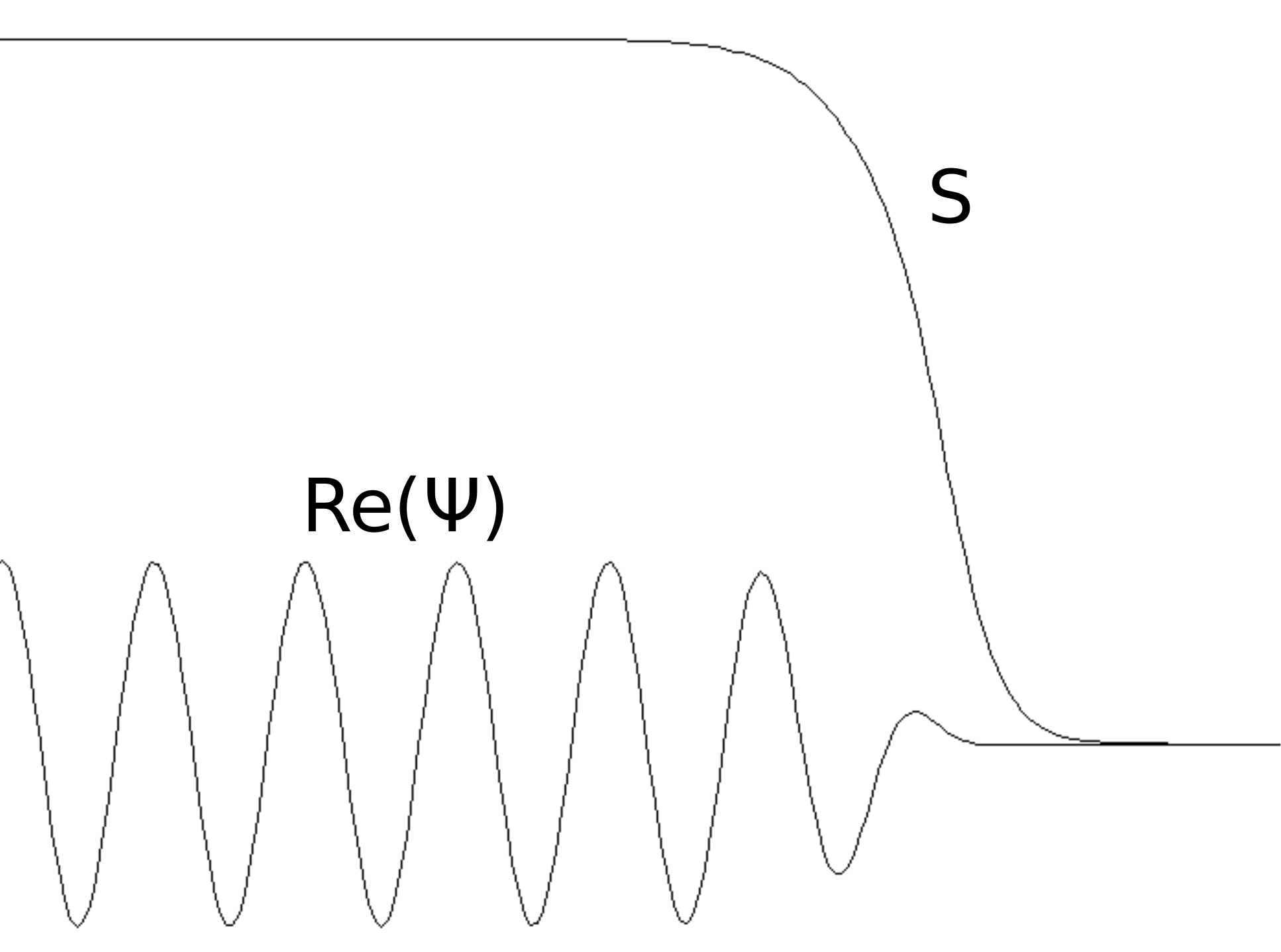}
\caption{Schematic of the one-dimensional simulation domain where the initial smectic-A nucleus grows (from left to right) into the unstable isotropic phase. Both the density modulation ($Re(\Psi)$, solid line) and scalar nematic order parameter ($S$, dotted line) are shown. The scalar smectic-A order parameter $\psi$ can be calculated using the both components of the complex order parameter $\psi = \sqrt{A^2+B^2}$}
\label{fig:simschem}
\end{figure}

Computational limitations constrain simulations to one-dimension in order to resolve the multiple length scales of smectic-A ordering ($nm$) and thermal diffusion ($\mu m$). Due to the use of the full complex order parameter (eq. \ref{eqsmec_order_param}) an additional constraint imposed on one-dimensional simulation requires that the smectic-A layers are parallel to the interface (imposed via initial/boundary conditions). A growing front where the smectic layers are perpendicular to the interface is inherently two-dimensional and cannot be captured using one-dimensional simulation.

\section{RESULTS AND DISCUSSION}

Simulations were performed for three different cases: shallow isothermal quench (neglecting the thermal energy balance, $T=331K$), shallow non-isothermal quench ($T=331K$), and deep non-isothermal quench ($T=330K$) where meta-stable nematic pre-ordering is predicted to be present (the timescales of nematic and smectic-A ordering differ greatly \cite{Abukhdeir2009a}). Shallow quench conditions describe the situation where the domain is quenched below the bulk isotropic/smectic-A transition temperature $T_b$ but above the lower stability limit of the isotropic phase $T_{AI}$. Subsequently, deep quench conditions describe a quench temperature below both $T_b$ and $T_{AI}$.

As previously mentioned, the vast majority of studies of diffusive dynamics of liquid crystalline materials is focused on mass transport. However, past experimental work by Dierking et al on growth laws of different types of pure-component liquid crystal mesophases \cite{Dierking2001,Bronnikov2004,Bronnikov2005,Dierking2003}, including a study of the direct isotropic/smectic-A transition \cite{Dierking2003}. The culmination of this work, following the findings of Huisman and Fassolino \cite{Huisman2007}, includes a generalized non-isothermal model \cite{Chan2008} showing the growth law evolution as a function of quench depth. This evolution transitions from purely non-conserved growth dynamics $L \propto t$ at deep quenches/undercooling to thermal diffusion-limited growth $L \propto t^{1/2}$ under shallow quench/undercooling conditions. The presented extension eqn \ref{eqn:final_energy_balance} of the high-order model of Mukherjee, Pleiner, and Brand \cite{deGennes1995,Mukherjee2001} to account for latent heat effects allows for the resolution of the minimum physics (\ref{fig:summary}) to describe the direct isotropic/smectic-A transition ofa pure-component liquid crystal.

\begin{figure*} 
\centering
\subfigure[]{\includegraphics[width=3in]{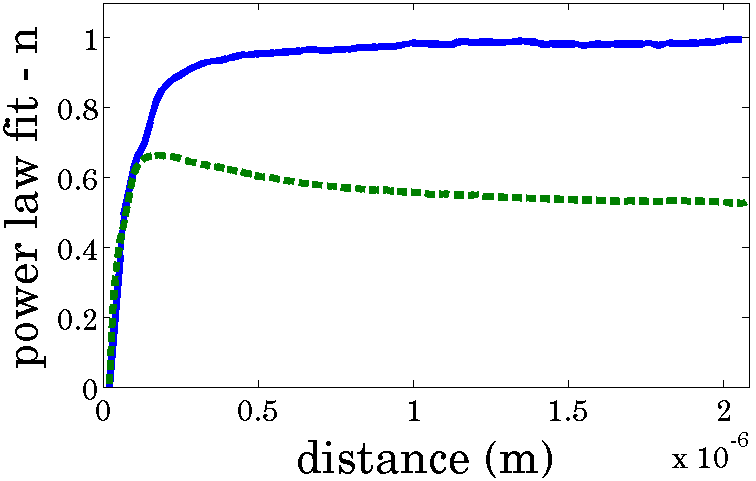}}
\subfigure[]{\includegraphics[width=3in]{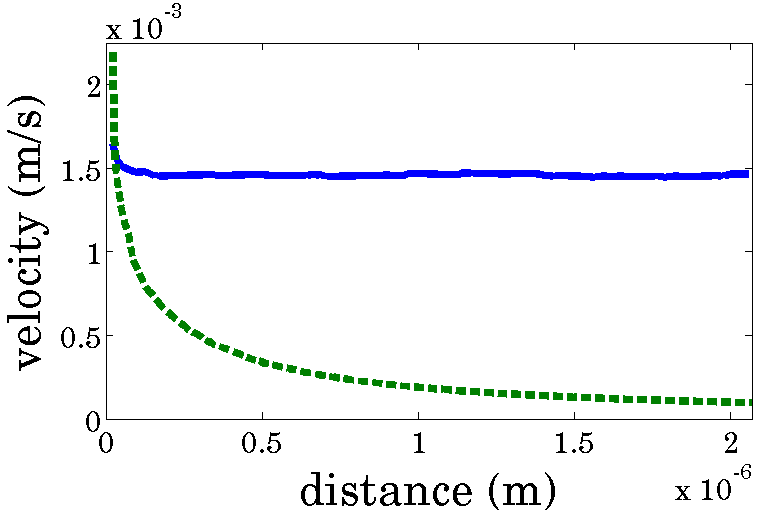}}
\caption{Plots of the a) power law fit  exponent ($n$ in $l \propto t^n$) and b) front velocity versus interface position for both the isothermal (solid line, $T=331K$ and $\mu_s/\mu_n = 25$) and non-isothermal (dashed line, $T=331K$ and $\mu_s/\mu_n = 25$) simulations. The material parameters and phenomenological coefficients, based upon 12CB \cite{Abukhdeir2008c}, are $T_{NI}=322.85K$, $T_{AI}=330.5K$, $a_0= 2\times10^5\frac{J}{m^3 K}$, $b=2.823\times10^7\frac{J}{m^3}$, $c=1.972\times10^7\frac{J}{m^3}$, $\alpha_0=1.903\times10^6\frac{J}{m^3 K}$, $\beta=3.956\times10^8\frac{J}{m^3}$, $\delta=9.792\times10^6 \frac{J}{m^3}$, $e=1.938\times10^{-11}pN$, $l_1=1\times10^{-12}pN$,$l_2=1.033\times10^{-12}pN$, $b_1=1\times 10^{-12}pN$, $b_2=3.334\times10^{-30}Jm$, and $\mu_n = 8.4\times10^{-2}\frac{N \times s}{m^2}$. Thermal material parameters used are based on values for 9CB and 10CB \cite{Pestov2003} (12CB data not available, $T_{ref}=330.0K$ for thermal conductivity values) $k_{\parallel} = 0.3100 \frac{W}{m K}$, $k_{\perp} = 0.1300 \frac{W}{m K}$, $C_p = 2600 \frac{J}{kg K}$, and $\rho = 1000 \frac{kg}{m^3}$.}
\label{fig:sim1}
\end{figure*}

\ref{fig:sim1}a-b shows the computed evolution of the power law exponent fit ($l \propto t^n$) and interface velocities for the first two simulation cases (isothermal and non-isothermal) versus position of the growing interface. \ref{fig:sim1}a shows that the inclusion of latent heat effects corrects the standard Landau-de Gennes model prediction of constant growth to that consistent with experimental observations of this specific system \cite{Dierking2003} and other liquid crystal mesophases \cite{Dierking2001,Bronnikov2004,Bronnikov2005,Dierking2003}. \ref{fig:sim1}b shows that in addition to the correction to the growth law, the trend and magnitude of the interface velocity is modified. The initial relative increase in the front velocity is due to the fact that in this work nucleation effects are neglected and the initial nucleus size and degree of liquid crystalline ordering are assumed to be at bulk equilibrium values (see Section \ref{sec:sim}). As the stable smectic-A nucleus grows into the unstable isotropic phase, there is an initial decrease in the degree of smectic-A order as predicted by the high-order model. Due to the inclusion of latent heat effects this decrease in smectic-A ordering requires thermal energy, effectively resulting in local cooling. The interface velocity can be approximated by \cite{Sutton1995}:
\begin{equation}
\beta w = \Delta F \left( T_s - T^* \right) - C
\end{equation}
where $\beta$ is the interfacial viscosity, $\Delta F \left( T_s - T^* \right)$ is the free energy driving force (free energy difference between the ordered/disordered phases) and $C$ is the capillary force. $T_s$ and $T^*$ are the interface and transition temperature, respectively. Thus any local cooling/heating has a subsequent effect of increasing/decreasing the interface velocity. 

\begin{figure*} 
\centering
\subfigure[]{\includegraphics[width=3in]{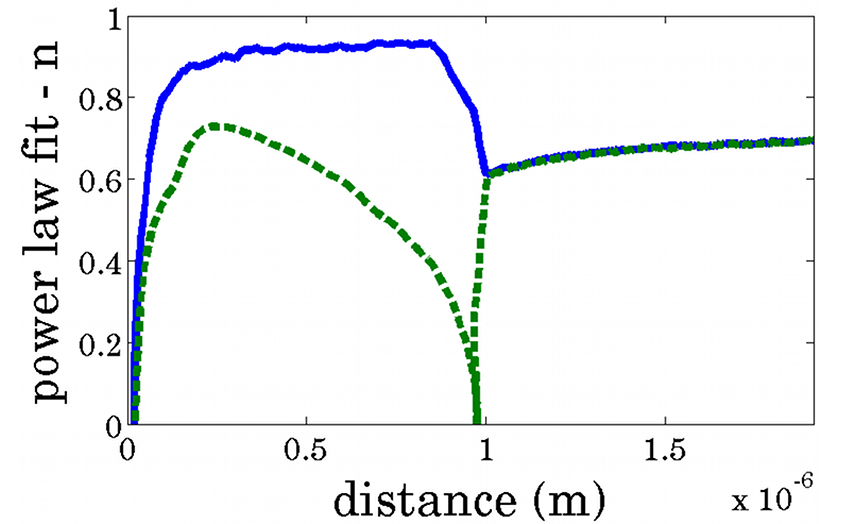}}
\subfigure[]{\includegraphics[width=3in]{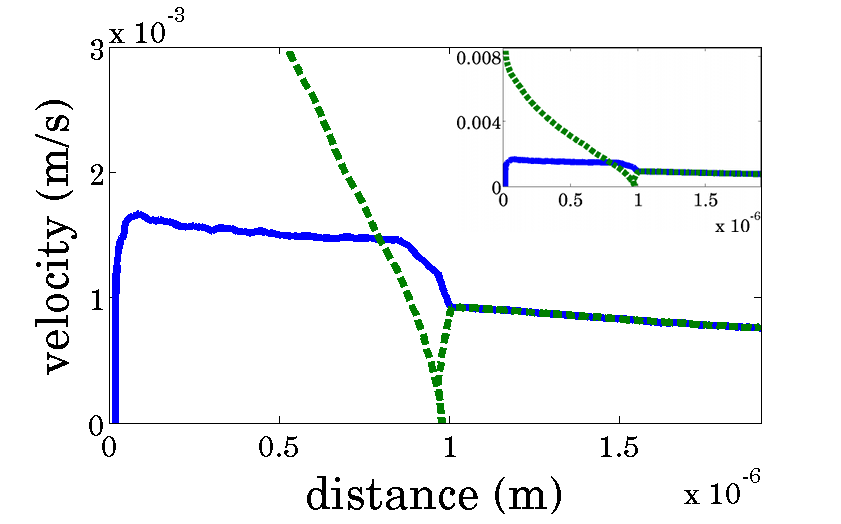}}
\caption{Plots of the a) power law fit exponent ($n$ in $l \propto t^n$) and b) front velocity versus interface position for the non-isothermal simulation where meta-stable nematic pre-ordering is present (solid line for the smectic-A front and dashed line for the nematic front, $T=330K$ and $\mu_s/\mu_n = 250$). The inset of b) shows the full scale of the nematic front velocity.}
\label{fig:sim2}
\end{figure*}

\ref{fig:sim2}a-b shows the evolution of the power law exponent fit ($l \propto t^n$) and interface velocities for the third simulation case (non-isothermal with meta-stable nematic pre-ordering) versus position of the growing interface. Under conditions where meta-stable nematic pre-ordering occurs, the effect of latent heat results in complex phase-ordering dynamics observed in other material systems \cite{Soule2009}.

In the present case, under deep quench/undercooling conditions, volume-driven growth is expected for the smectic-A phase. Conversely, any meta-stable nematic pre-ordering is under effectively shallow quench conditions (see Figure 2a of ref. \cite{Abukhdeir2009a} for the free energy landscape). As is seen in \ref{fig:sim2}a, the smectic-A front initially undergoes constant volume-driven growth preceded by a meta-stable nematic front which rapidly exhibits diffusion-limited dynamics. As the meta-stable nematic front velocity decreases (\ref{fig:sim2}b) it is overtaken by the smectic-A front. This non-monotonic acceleration/deceleration/acceleration behavior is similar to dual phase-ordering/chemical-demixing growth kinetics recently simulated in nematic PDLC systems \cite{Soule2009}. A unique differentiation in the growth dynamics as compared to those observed in the phase-ordering/chemical-demixing system is that the evolution of latent heat from the trailing smectic-A front contributes to the isotropic/metastable nematic interface temperature which results in a small time period where the meta-stable nematic front begins to melt/recede (shown in \ref{fig:sim2}b).

\begin{figure} 
\centering
\includegraphics[width=2.5in]{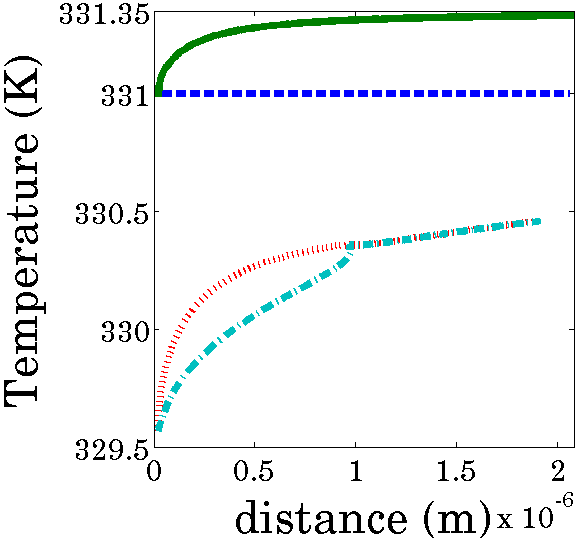}
\caption{Plot of the interfacial temperature versus interface position of the three different simulations: non-isothermal (solid line), isothermal (dashed line), and non-isothermal with meta-stable nematic pre-ordering (stippled line for the smectic-A front and dash-dot line for the nematic front).}
\label{fig:int_temp}
\end{figure}

\ref{fig:int_temp} shows the interface temperature versus position of the growing interface for all three simulation cases. Inclusion of the latent heat contribution from phase-ordering causes the interfacial temperature to approach the bulk transition temperature $T_b$, which subsequently causes the free energy driving force of the isotropic/smectic-A transition to approach zero. Thus, under these conditions, thermal diffusion limits growth resulting in dynamics converging to a power law exponent of $n=\frac{1}{2}$ as observed experimentally \cite{Dierking2001,Bronnikov2004,Bronnikov2005,Dierking2003} and predicted theoretically \cite{Huisman2007,Chan2008,Abukhdeir2008b} for liquid crystal growth under shallow quench/undercooling conditions. In the case of metastable nematic pre-ordering, the evolution of the isotropic/metastable nematic front temperature reveals the influence of the latent heat contribution of the trailing smectic-A front. The thermal energy from the additional heat source causes non-monotonic behavior of the meta-nematic interface temperature evolution, favoring the eventual merging of the two fronts. Following this merging event, convergence of the isotropic/smectic-A front velocity is not accessible due to computational limitations. The interface velocity is predicted to converge to a constant value due to the deep quench conditions and assumption of infinite volume (see Section \ref{sec:sim}). These interesting dynamics suggest the possibility of oscillatory splitting/merging behavior of systems exhibiting metastable nematic preordering under realistic experimental conditions where ideal undercooling is not feasible.

\section{CONCLUSION}

A thermal energy balance was derived to extend a high-order Landau-de Gennes type model of the isotropic/smectic-A liquid crystalline transition to take into account latent heat of phase-ordering, anisotropic thermal diffusion, and dissipation (eqn. \ref{eqn:final_energy_balance}). One-dimensional simulations were performed showing that the inclusion of the thermal energy balance corrects the standard Landau-de Gennes predictions of volume-driven growth under shallow quench conditions (\ref{fig:sim1}a-b) to experimentally observed \cite{Dierking2001,Dierking2003} diffusion-limited growth. Additionally, the effect of incorporating thermal effects in the presence of meta-stable nematic pre-ordering was found to result in non-monotonic acceleration/deceleration/acceleration growth dynamics (\ref{fig:sim2}a-b), as observed experimentally \cite{Tokita2006}. This work sets the basis for further study of non-isothermal effects of the direct isotropic/smectic-A transition where an extended subset of liquid crystal physics present in the transformation is included (\ref{fig:summary}). The presented extension to the high-order model of Mukherjee, Pleiner, and Brand \cite{deGennes1995,Mukherjee2001} allows for full three-dimensional multi-scale multi-transport simulation incorporating the diverse defect and texturing dynamics of lamellar smectic-A ordering coupled with non-negligible thermal and dissipative effects.

\section*{Acknowledgment}

This work was supported by a grant from the Natural Science and Engineering Research Council of Canada.

\bibliographystyle{unsrt}
\bibliography{/home/nasser/Documents/references/references}

\end{document}